\documentclass[a4paper,onecolumn,11pt,accepted=2025-03-22]{quantumarticle}
\pdfoutput=1
\usepackage[utf8]{inputenc}
\usepackage[english]{babel}
\usepackage[T1]{fontenc}
\usepackage{amsmath}
\usepackage{hyperref}
\usepackage{color}

\usepackage{tikz}
\usepackage{lipsum}

\usepackage[sort]{cite}

\begin{document}

\title{Mpemba effect and super-accelerated thermalization in the  damped quantum harmonic oscillator}

\author{Stefano Longhi}
\affiliation{Dipartimento di Fisica, Politecnico di Milano, Piazza L. da Vinci 32, I-20133 Milano, Italy \&
IFISC (UIB-CSIC), Instituto de Fisica Interdisciplinar y Sistemas Complejos - Palma de Mallorca, Spain}
\orcid{0000-0002-8739-3542}
\email{stefano.longhi@polimi.it}

\maketitle

\begin{abstract}
  The behavior of systems far from equilibrium is often complex and unpredictable, challenging and sometimes overturning the physical intuition derived from equilibrium scenarios. One striking example of this is the Mpemba effect, which implies that non-equilibrium states can sometimes relax more rapidly when they are further from equilibrium. Despite a rich historical background, the precise conditions and mechanisms behind this phenomenon remain unclear. Recently, there has been growing interest in investigating accelerated relaxation and Mpemba-like effects within quantum systems. In this work, 
we explore a quantum manifestation of the Mpemba effect in a simple and paradigmatic model of open quantum systems: the damped quantum harmonic oscillator, which describes the relaxation of a bosonic mode in contact with a thermal bath at finite temperature $T$. By means of an exact analytical analysis of the relaxation dynamics {based on  the method of moments in both population and coherence subspaces}, we demonstrate that any initial distribution of populations with the first $r$ moments exactly matching those of the equilibrium distribution shows a super-accelerated relaxation to equilibrium at a rate linearly increasing with $r$, leading to a pronounced Mpemba effect. In particular, one can find a broad class of far-from-equilibrium distributions that relax to equilibrium faster than any other initial thermal state with a temperature $T'$ arbitrarily close to $T$. {The super-accelerated relaxation effect is shown to persist even for a broad class of initial states with non-vanishing coherences, and a general criterion for the observation of super-accelerated thermalization is presented.}
\end{abstract}

\section{Introduction}
The Mpemba effect is a fascinating and counterintuitive phenomenon in which a warmer system cools more rapidly than a cooler one when both are placed in the same environment and allowed to reach thermal equilibrium. First documented in the context of water freezing \cite{M1,M2,M3}, the effect has sparked significant curiosity and debate within the scientific community, inspiring efforts to identify and understand its underlying mechanisms \cite{M4,M5,M6,M7,M8,M9,M10,M11,M12,M12b}. While numerous explanations have been proposed, ranging from non-equilibrium thermodynamics to variations in heat transfer rates, the Mpemba effect remains a subject of active research \cite{M10}.
In recent years, quantum versions of the classical thermal Mpemba effect have attracted considerable attention \cite{QM1,QM2,QM3,QM4,QM5,QM6,QM7,QM8,QM9,QM10,QM11,QM12,QM13,QM14,QM15,QM16,QM17,QM18,QM19,QM20,QM21,QM22,QM23,QMa}. The exploration of quantum Mpemba-like phenomena has opened new avenues to understand thermalization processes in quantum systems, where distinct and often non trivial behaviors emerge due to quantum coherence and dissipation. This quantum perspective not only enriches our comprehension of thermalization dynamics but also raises fundamental questions about the interplay between quantum effects and classical thermodynamics \cite{QM24}.

In this work, we investigate the manifestation of the Mpemba effect in the thermalization process of a damped quantum harmonic oscillator, i.e. in the relaxation dynamics of a bosonic mode in contact with a thermal reservoir \cite{L1,L2,L3,dec1}. The dissipative quantum harmonic oscillator provides a simple and ubiquitous model in the physics of open quantum systems, frequently used to describe a wide range of quantum systems, from trapped ions and nanomechanical resonators to the quantized modes of electromagnetic field.  When coupled to an external thermal reservoir, the dynamics of a quantum harmonic oscillator exhibit interesting relaxation behavior, which can reveal intriguing non-equilibrium phenomena such as decoherence and quantum to classical transition \cite{dec1,dec2,dec3,dec4,dec5,dec6,dec7}. The damped quantum oscillator also describes the photon statistics properties of a laser below threshold \cite{Scully,Gen1,Gen2,uffa}.
Despite its simplicity and a plethora of research, the occurrence of the Mpemba effect in this foundational model has largely been overlooked. Recently, a quantum Mpemba effect for bosonic modes in contact with a zero-temperature bath, exploiting quantum states of light, has been predicted \cite{QM22}. {Here we unveil super-accelerated relaxation and the occurrence of the Mpemba effect in the damped quantum harmonic oscillator in contact with a bath of finite temperature $T$. To this aim, relaxation dynamics in both population and coherence subspaces is analytically investigated by spectral methods and using the method of moments, which can elegantly provide general conditions for super-accelerated thermalization.}
We demonstrate that any initial distribution of populations, whose first $r$ moments match those of the equilibrium distribution, displays an accelerated relaxation to equilibrium at a rate that increases linearly with $r$, leading to a pronounced Mpemba effect. In particular, there exist very far from equilibrium  distributions that relax to equilibrium faster than any other initial thermal state with a temperature $T'$ arbitrarily close to $T$. {Accelerated relaxation can also arise for a broad class of initial states with non-vanishing coherences.}
 Our findings reveal the presence of the Mpemba effect and accelerated thermalization within a foundational and ubiquitous quantum model, offering insights that could be valuable for quantum technologies where rapid relaxation dynamics are advantageous.

\section{Damped quantum harmonic oscillator}
The starting point of our analysis is provided by a widely-studied model of a quantum harmonic oscillator (a bosonic mode) at frequency $\omega_0$ weakly coupled to a reservoir in thermal equilibrium at temperature $T$ \cite{
L1,L2,L3,dec1,dec2,dec5,dec6,dec7,O1,O2,O3}.  In the Born and Markov approximations, the reduced density operator $\hat{\rho}(t)$ for the harmonic oscillator {in the Schr\"odinger picture} obeys the following  quantum master equation (see e.g. \cite{L1})
\begin{equation}
\frac{d \hat{\rho}}{dt}=-i[ \hat{H}, \hat{\rho}]+ \gamma(n_{th}+1) ( 2 \hat{a} \rho \hat{a}^{\dag}-\hat{a}^{\dag} \hat{a} \hat{ \rho} - \hat{\rho} \hat{a}^{\dag} \hat{a}) + \gamma n_{th} (2 \hat{a}^{\dag} \hat{\rho} \hat{a} - \hat{a} \hat{a}^\dag \hat{\rho}- \hat{\rho} \hat{a} \hat{a}^\dag) \equiv \mathcal{L} \hat{\rho}
\end{equation}
where { $\hat{H}= \omega_0 \hat{a}^{\dag} \hat{a}$ is the Hamiltonian of the quantum oscillator, $\mathcal{L}$ is the Lindbladian}, $\hat{a}$ and $\hat{a}^\dag$ are the usual bosonic annihilation and creation operators, $\gamma$ is the damping rate, and
\begin{equation}
n_{th} = \frac{1}{\exp(\hbar \omega_0 /k_BT)-1}
\end{equation}
is the mean number of quanta in a mode with energy $\hbar \omega_0$ of the thermal reservoir. {This model belongs to the general class of open quantum systems coupled to a Markovian heat bath,  described by a Davies map, where a general condition for accelerated thermalization has been derived in a recent work \cite{QM16}. For an initial state $\hat{\rho}_i$, the general evolution of the system reads
\begin{equation}
\hat{\rho}(t)=\exp( \mathcal{L} t) \hat{\rho}_i.
\end{equation}
Since $\exp( \mathcal{L} t ) $ is block diagonal in the energy (Fock) eigenbasis $|n \rangle= (1/ \sqrt{n!}) a^{\dag n} |0 \rangle$, we can
write $\mathcal{L}=\mathcal{L}_P \bigoplus\mathcal{L}_C$,  
where $\mathcal{L}_P$ is the population sub-block,
whose right eigenmatrices are diagonal in the energy
eigenbasis, and $\mathcal{L}_C$ is the coherence sub-block.}
In fact, the evolution equations for the diagonal elements $P_n(t)=\rho_{n,n}(t)= \langle n | \rho(t) |n \rangle$ of the density matrix (populations) are decoupled from those of coherences $\rho_{n,m}= \langle n | \rho(t) |m \rangle$ ($n \neq m$), namely one has
\begin{equation}
\frac{dP_n}{dt}=-2 \gamma \left[ n(2 n_{th}+1)+n_{th} \right]P_n+2 \gamma (n_{th}+1)(n+1)P_{n+1}+2 \gamma n n_{th}P_{n-1}
\end{equation}
for the populations ($n=0,1,2,3,...$), and
\begin{eqnarray}
\frac{d \rho_{n,m}}{dt} & = &  2 \gamma \sqrt{(n+1)(m+1)} (1+{n}_{th}) \rho_{n+1,m+1}+2 \gamma {n}_{th} \sqrt{nm} \rho_{n-1,m-1}  \nonumber \\
 & - &  \gamma [(n+m)(1+2 {n}_{th})+2{n}_{th}] \rho_{n,m} +i(m-n) \omega_0 \rho_{n,m}
\end{eqnarray}
for the coherences ($n,m=0,1,2,3,...$, $n \neq m$). In the long time limit $t \rightarrow \infty$, the solution $\hat{\rho}(t)$ relaxes toward the stationary state $\hat{\rho}^{(S)}$, which is given by the thermal distribution (see e.g. \cite{L1})
\begin{equation}
P_n^{(S)}=\frac{1}{1+n_{th}} \left( \frac{n_{th}}{1+n_{th}} \right)^n=\left[  1- \exp(-\hbar \omega_0/k_BT) \right] \exp \left( - \frac{n \hbar \omega_0}{k_B T} \right)
\end{equation}
for populations and $\rho_{n,m}^{(S)}=0$ for coherences ($n \neq m$). The average number of excitation at thermal equilibrium is clearly $\langle n \rangle ={\rm Tr}( \hat{\rho}^{(S)} \hat{a}^\dag \hat{a})=n_{th}$, which follows the Planck distribution Eq.(2). {The relaxation dynamics of $\hat{\rho}(t)$ toward the stationary state is governed by the eigenvalues $\lambda_k$ and corresponding left and right eigenoperators, $\hat{l}_k$ and $\hat{r}_k$, of the Lindbladian $\mathcal{L}=\mathcal{L}_P \bigoplus\mathcal{L}_C$ via the relation \cite{QM3,QM16}
\begin{equation}
\hat{\rho}(t)=\hat{\rho}^{(S)}+ \sum_k {\rm Tr} \left(  \hat{l}_k \hat{\rho}_i \right) \hat{r}_k \exp( \lambda_k t)
\end{equation}
where the index $k$ spans  the entire eigenvalues in both the population ($\mathcal{L}_P$) and coherence ($\mathcal{L}_C$) subspaces, excluding $\lambda_k=0$ corresponding to the stationary state $\hat{\rho}^{(S)}$. The speed of the relaxation toward equilibrium is determined by the decay rate $-{\rm Re}(\lambda_k)$ of the slowest decaying eigenoperator. For the damped quantum harmonic oscillator, the coherence sub-block $\mathcal{L}_C$ can be further decomposed as  $\mathcal{L}_C=\bigoplus_{s= \pm1, \pm2, \pm3,...} \mathcal{L}_C^{(s)}$, where $\mathcal{L}_C^{(s)}$ is the sub-block of the Lindbladian that describes the relaxation of coherences $\rho_{n,n+s}(t)$ with fixed $s$. The full set of eigenvalues $\lambda_k$ can be calculated analytically and reads [Fig.1(a)]
\[
\lambda_k=-2 \gamma(\alpha+|s|/2)+i \omega_0 s 
\]
where $k=(\alpha,s)$, $ \alpha=0,1,2,3,...$ and $s=0, \pm 1 , \pm 2 ,...$. The eigenvalues of the population subspace correspond to the subset $s=0$, whereas the eigenvalues of the coherence subspace are those with $s \neq 0$.  The slowest decaying eigenoperator  belongs to the coherence subspace ($s= \pm 1$), with a decay rate $\gamma$. As suggested in Ref.\cite{QM16}, when the slowest decaying eigenoperator belongs to the coherence subspace, starting from an initial state with coherences in the energy eigenbasis
 an exponential speedup to equilibrium will always occur if the state is first rotated, via a unitary transformation ${U}$, to a
diagonal state in the energy eigenbasis. This general result suggests us to focus our attention to the population relaxation dynamics subspace: as we will show in Sec.3, whenever the initial state $\hat{\rho}_i$ is diagonal in the energy basis and the first $r$ moments of the distribution $(\rho_i)_{n,n}$ match those of the thermal (equilibrium) distribution, a super-accelerated thermalization occurs, which results in a pronounced Mpemba effect. As discussed in the last section (Sec.4), such a super-accelerated relaxation persists for a broader class of initial states with coherences, without the need to first perform a unitary transformation to a diagonal  state.}



\begin{figure}[h]
\includegraphics[width=15cm]{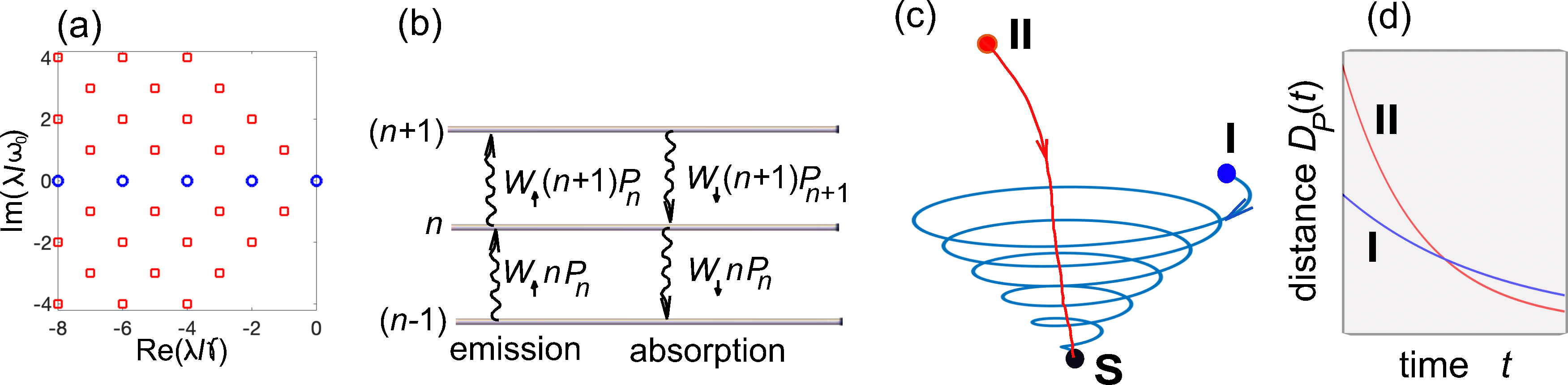}
\caption{(Color online) {(a) Spectrum $\lambda$ of the Lindbladian $\mathcal{L}={\mathcal L}_P \bigoplus \mathcal{L}_C$ of the damped quantum harmonic oscillator. The blue circles are the eigenvalues of the population sub-block $\mathcal{L}_P$, whereas the red squares are the eigenvalues of the coherence sub-block $\mathcal{L}_C=\bigoplus_s \mathcal{L}^{(s)}_C$, with $s= \pm1, \pm 2 , \pm3,...$. The $\lambda=0$ eigenvalue corresponds to the equilibrium thermal distribution S.}
 (b) Schematic of the absorption and emission of quanta of the quantum harmonic oscillator in contact with a thermal bath at temperature $T$, as described by the Pauli master equation (8).The transition rates $W_{\uparrow}$ and $W_{\downarrow}$ are given by Eq.(9). (c) Cartoon of the Mpemba effect. An initial far-from equilibrium population distribution II can relax toward the equilibrium thermal state S at temperature $T$ faster than a closer-to-equilibrium initial population distribution I. {The two curves, connecting states I and II with S, schematically depict the phase space trajectories toward equilibration.} (d) Schematic of the relaxation dynamics (distance from the equilibrium state S versus time $t$) for the population distributions I and II. Thermalization of population distribution II is faster because it does not excite the slow decaying mode of the Markov transition matrix $M$. Super-accelerated thermalization occurs when the initial population distribution does not excite the first $r$ decaying modes, with $r>1$. This happens when the initial out-of equilibrium distribution of populations matches the first $r$ moments of the equilibrium (thermal) distribution.}
\end{figure}

\section{Super-accelerated {population} relaxation dynamics and the Mpemba effect}
\subsection{Population dynamics: Pauli master equation}
{To unveil accelerated thermalization and the emergence of the Mpemba effect in the relaxation of the quantum oscillator, as suggested in \cite{QM16}  let us assume that the initial state $\hat{\rho}_i$ is initially transformed, via a unitary evolution ${U}$, to a state $U \hat{\rho}_i$ which is diagonal in the energy basis. The subsequent relaxation dynamics thus occurs in the population subspace $\mathcal{L}_P$ and thermalization} is described by the Pauli master equation (4) {for the population $P_n(t)$ describing the distribution of excitation quanta in the oscillator}. This equation can be written in the useful form
\begin{equation}
\frac{dP_n}{dt}=W_{\downarrow}(n+1)P_{n+1}+W_{\uparrow}n P_{n-1}-W_{\uparrow}(n+1)P_n-W_{\downarrow}nP_n \equiv \sum_l M_{n,l} P_l
\end{equation}
($n=0,1,2,3,...$), where we have set
\begin{equation}
W_{\uparrow}= 2 \gamma n_{th}  \; ,\;\;\; W_{\downarrow}=2 \gamma (n_{th}+1)=W_{\uparrow} \exp(\hbar \omega_0 / k_BT)
\end{equation}
This is a classical rate equation describing the change in populations due to absorption and emission of quanta with the reservoir at rates $W_{\uparrow}$ and $W_{\downarrow}$,  as schematically shown  in Fig.1(b). 
The thermalization process of the populations is entirely captured by the spectral properties of the Markov transition matrix $M$ entering in the rate equation (8), which is a tridiagonal (Jacobi) semi-infinite matrix. The spectral analysis of matrix $M$ is presented in subsection 3.2 and Appendix A.1. Major insights into the relaxation process are also gained by considering the dynamical 
evolution of the moments associated to the population distribution $P_n(t)$, which is presented in the subsection 3.3 and in the Appendix A.2. {The method of moments is particularly elegant and powerful since it can provide the conditions for super-accelerated relaxation without any explicit determination of the eigenvalues and corresponding left/right eigenoperators of $\mathcal{L}_P$.}\\
 
\subsection{Spectral analysis}
 Let us indicate by $\lambda_{\alpha} \leq 0$ the eigenvalues and by $\psi_n^{(\alpha)}$ and $\phi_n^{(\alpha)}$ the corresponding right and left eigenvectors of $M$, respectively. These are defined by the eigenvalue equations 
\begin{equation}
\sum_{l=0}^{\infty} M_{n,l} \psi_l^{(\alpha)} =\lambda_{\alpha} \psi_n^{(\alpha)}
\end{equation}
for the right eigenvectors (or just eigenvectors), and
\begin{equation}
\sum_{l=0}^{\infty} M^{\dag}_{n,l} \phi_l^{(\alpha)} =\lambda_{\alpha} \phi_n^{(\alpha)}
\end{equation}
for the left eigenvectors ($\alpha=0,1,2,3,...$), where $M^{\dag}$ is the adjoint of $M$, i.e. $M^{\dag}_{n,l}=M^*_{l,n}=M_{l,n}$. 
{We note that, in operatorial form, $\hat{l}_{\alpha}=\phi_n^{(\alpha)} |n \rangle \langle n|$ and $\hat{r}_{\alpha}=\psi_n^{(\alpha)} |n \rangle \langle n|$ are the left and right eigenoperators of $\mathcal{L}_P$, respectively, with eigenvalue $\lambda_{\alpha}$.}
Clearly, $\psi_l^{(\alpha)}$ and $\phi_l^{(\alpha)}$ can be assumed to be real, and normalization is taken such that the orthonormality condition 
\[ \sum_n \phi_n^{(\alpha)} \psi_n^{(\beta)}= \delta_{\alpha,\beta} \] 
holds. Owing to the form of the Jacobi matrix $M$ \footnote{The Jacobi matrix $M$ is not Hermitian, however it can be transformed into a Hermitian symmetric matrix $M'$ via a non-unitary transformation $M^{\prime}=U M U^{-1}$ with $U_{n,m}= \exp[(n/2) \hbar \omega_0/k_BT)] \delta_{n,m}$. This justifies the simple relation Eq.(10) between the eigenfunctions $\psi^{(\alpha)}_n$ of $M$ and $\phi_n^{(\alpha)}$ of the adjoint $M^{\dag}$.}, it can be readily shown that the following simple relation can be established between left and right eigenvectors
\begin{equation}
\phi_n^{(\alpha)}= \mathcal{B}_{\alpha} \exp\left( \frac{n \hbar \omega_0}{k_BT} \right) \psi_n^{(\alpha)}
\end{equation}
where $\mathcal{B}_{\alpha}$ is a suitable normalization constant.
Since the Jacobi matrix $M$ is semi-infinite, care should be taken in the definition of eigenvalues and eigenvectors, because they depend not only on $M$ but also on the Hilbert space $\mathcal{H}$, i.e. the set of sequences $\{ \psi_n\}_{n=0}^{\infty}$, to which the matrix $M$ acts (see e.g. \cite{matematico}). Here we assume for the Hilbert space $\mathcal{H}$ the sequences $\{ \psi_n \}$ such that $\sum_{n=0}^{\infty} |\psi_n| \exp(i \sigma^+ n) < \infty$, where $\sigma=0^+$ is an arbitrarily small positive number. As shown in the Appendix A.1, in the Hilbert space $\mathcal{H}$  the spectrum of $M$ is pure point and  
  the discrete set of eigenvalues $\lambda_{\alpha}$ are real non-positive numbers, which can be ordered such that 
\[
\lambda_0=0>\lambda_1>\lambda_2>\lambda_3>..... \]
 The zero eigenvalue $\lambda_0=0$ corresponds to the stationary (thermal) state $P_n^{(S)}$, given by Eq.(6), i.e. one has
\begin{equation}
\psi_n^{(0)}=P_n^{(S)}=\frac{1}{1+n_{th}} \left( \frac{n_{th}}{1+n_{th}} \right)^n=\left[  1- \exp(-\hbar \omega_0/k_BT) \right] \exp \left( - \frac{n \hbar \omega_0}{k_B T} \right).
\end{equation}
The corresponding left eigenstate is the extended state $\phi_n^{(0)}=1$, as it readily follows from Eqs.(12) and (13).
The analytic form of all other eigenvalues $\lambda_{\alpha}$ and corresponding (right) eigenvectors $\psi_n^{(\alpha)}$  is derived in the Appendix A.1. Here we just provide some of the main steps to solve the spectral equation
\begin{equation}
E \psi_n=- \left[ n(2 n_{th}+1)+n_{th} \right] \psi_n+(n_{th}+1)(n+1)\psi_{n+1}+ n n_{th} \psi_n
\end{equation}
where $E= \lambda/(2 \gamma)$ is the eigenvalue in units of $2 \gamma$. First, one introduces the generating function
\begin{equation}
G(z)=\sum_{n=0}^{\infty} \psi_n z^n
 \end{equation}
 for the eigenfunction $\psi_n$, where $z$ is a complex number with $|z|<R_0$ and $R_0>1$ is the radius of convergence of the series. The eigenfunction $\psi_n$ is obtained from the inversion relation
 \begin{equation}
 \psi_n=\frac{1}{2 \pi i} \oint_{\mathcal{C}} dz G(z) z^{-n-1},
 \end{equation}
 where $\mathcal{C}$ is a closed loop in complex plane encircling the origin $z=0$ and internal to the convergence circle. 
 The generating function $G(z)$ is then found to satisfy the following differential equation
 \begin{equation}
 \frac{1}{G} \frac{dG}{dz}=\frac{E+n_{th}-n_{th} z}{n_{th}z^2-(1+2n_{th})z+1+n_{th}} \equiv F(z)
 \end{equation}
 in complex $z$ plane. Since $G(z)$ is holomorphic inside the convergence circle, the condition
 \begin{equation}
 \oint_{\mathcal{C}} dz \frac{1}{G} \frac{dG}{dz}= 2 \pi i \alpha,
 \end{equation}
 must be satisfied, i.e.
 \begin{equation}
 \oint_{\mathcal{C}} dz F(z)=  2 \pi i \alpha,
 \end{equation}
 where $\alpha$ is an integer number. This 'quantization' condition enables to determine the eigenvalues $E$, after computing the integral on the left hand side of Eq.(19) with the residue theorem and using for $F(z)$ the expression defined by Eq.(17). This yields $E=-\alpha$. i.e. the eigenvalues $\lambda_{\alpha}$ of $M$ are given by
 \begin{equation}
 \lambda_{\alpha}=-2 \gamma \alpha.
 \end{equation}
Since $\lambda_{\alpha}$ is non-positive, the integer $\alpha$ can take the values $\alpha=0,1,2,3,...$. Equation (20) provides the entire set of eigenvalues of the Lindbladian $\mathcal{L}_P$ in the population subspace. The expression of the corresponding eigenfunctions $\psi_n=\psi_n^{(\alpha)}$ are obtained from the inversion relation (16) and reads (technical details are given in the Appendix A.1)
\begin{equation}
 \psi_n^{(\alpha)}= \frac{1}{ n_{th} \alpha !} \left( \frac{ d^{\alpha}}{dz^{\alpha}} 
 \frac{(z-1)^{\alpha}}{z^{n+1}} \right)_{z=1+1/n_{th}}.
\end{equation}
 For $\alpha=0$, the right eigenfunctions $\psi_n^{(\alpha)}$ as given by Eq.(21) clearly yields the stationary (thermal) distribution $P_n^{(S)}$ Eq.(13). The right eigenfunction of the slowest decaying mode is obtained from Eq.(21) by letting $\alpha=1$ and reads explicitly
 \begin{equation}
 \psi_n^{(1)}= \frac{n_{th}}{1+n_{th}} \left(  1-\frac{n}{n_{th}} \right) P_n^{(S)}.
 \end{equation}
The corresponding left eigenfunction is obtained using Eq.(12), with $\mathcal{B}_1$ calculated from the normalization condition $\sum_{n=0}^{\infty} \phi_n^{(1)} \psi_n^{(1)}=1$, and reads
\begin{equation}
\phi_n^{(1)}=   1-\frac{n}{n_{th}}.
\end{equation}
{In the operatorial form, the left eigenoperator of the slowest decaying mode in the population subspace $\mathcal{L}_P$ can be written as $\hat{l}_1=\phi_n^{(1)}|n \rangle \langle n|=1-\hat{n} /n_{th}$, where $\hat{n}=\hat{a}^{\dag} \hat{a}$ is the number operator.} The eigenfunctions of higher orders can be calculated in a similar way. In particular, it turns out that the left eigenfunction $\phi_{n}^{(\alpha)}$  is a polynomial of degree $\alpha$, { and in operatorial form  $\hat{l}_{\alpha}$ is a polynomial of $\hat{n}$ of order $\alpha$.}

\subsection{Accelerated thermalization and the Mpemba effect}
Let us consider an initial population distribution $P_n(0)$ with $P_n(0) \rightarrow 0$ fast enough as $n \rightarrow \infty$, such that all moments of the distribution, $I_l \equiv \sum_{n=0}^{\infty}n^lP_n(0)$ ($l=1,2,3,...$) are finite \footnote{In case the distribution $P_n(0)$ does not have finite moments, one should consider the initially-perturbed population distribution $P_n(0) \exp(-\sigma n)$ and taking the limit $\sigma \rightarrow 0^+$. Since for such a distribution the mean number of excitation $\sum_{n=1}^{\infty} n P_n(0)$ diverges as $\sigma \rightarrow 0^+$, the condition given by Eq.(25) is never satisfied and the relaxation to equilibrium cannot be accelerated.}. In this case the solution to the master equation (4) can be written as
\begin{equation}
P_n(t)=\sum_{\alpha=0}^{\infty} C_{\alpha} \psi_n^{(\alpha)} \exp(\lambda_{\alpha}t)=P_n^{(S)}+\sum_{\alpha=1}^{\infty} C_{\alpha} \psi_n^{(\alpha)} \exp(\lambda_{\alpha}t)
\end{equation}
where the spectral amplitudes $C_{\alpha}$ are given by
{
\begin{equation}
C_{\alpha}={\rm Tr} \left(  \hat{l}_{\alpha} \hat{\rho}_i \right)= \sum_{n=0}^{\infty} P_n(0) \phi_n^{(\alpha)}
\end{equation}
}
with $C_0=\sum_n P_n(0)=1$. Note that, since all moments of $P_n(0)$ are finite and $\phi_n^{(\alpha)} \sim n^{\alpha}$ as $n \rightarrow \infty$, the spectral amplitudes $C_{\alpha}$ are not singular.
Equation (24) shows that, since $\lambda_{\alpha}=-2 \alpha \gamma$, as $t \rightarrow \infty$ the populations $P_n(t)$ approach the thermal (equilibrium) distribution $P_n^{(S)}$. 
The relaxation rate toward equilibrium is basically established by the slowest decaying mode ('slow mode') entering in the power series on the right hand side of Eq.(24), whose spectral amplitude $C_1$ is readily obtained from Eqs.(23), (25) and reads
\begin{equation}
C_1=1- \frac{1}{n_{th}}\sum_{n=0}^{\infty} nP_n(0).
\end{equation}
 For a rather generic initial state $P_n(0)$ such that $C_1 \neq 0$, the relaxation rate is given by $|\lambda_1|=2 \gamma$. However, for special initial states such that $C_1=0$, i.e. when the mean number of excitations of the distribution $P_n(0)$ is exactly $n_{th}$,
 \begin{equation}
 \sum_{n=0}^{\infty} nP_n(0)=n_{th},
 \end{equation}
  the relaxation is faster, with a rate no smaller than $| \lambda_2|=4 \gamma$.\\
   More generally, let us assume that the initial non-equilibrium population distribution $P_n(0)$ has all moments $I_l$, for $l=1,2,..,r$ equal to the ones of the equilibrium distribution $P_{n}^{(S)}$, i.e.
  \begin{equation}
  I_l=\sum_{n=0}^{\infty} n^l P_n(0)= \sum_{n=0}^{\infty} n^l P_n^{(S)}(0) 
  \end{equation}
($r$ is a positive integer), and $I_{r+1} \neq \sum_{n=0}^{\infty} n^{r+1}P_n^{(S)}(0)$. Then it can be proven that $C_{\alpha}=0$ for $\alpha=1,2,3,.., r$ and $C_{r+1} \neq 0$, so that the relaxation of $P_n(t)$ toward the equilibrium thermal distribution occurs at the rate $|\lambda_{r+1}|=2 \gamma(r+1)$. This result, corresponding to a super-accelerated thermalization for a large integer $r$, is demonstrated in an elegant way in the Appendix A.2 using the method of moments, where the dynamical equations for the moments $Q_l(t)=\sum_{n=1}^{\infty} n^lP_n(t)$ of the distribution $P_n(t)$ are derived. { This method is very powerful, as it allows one to establish the conditions for super-acceleration --i.e., the vanishing of the projection amplitudes $C_{\alpha}$ --without the need for explicitly determining the left eigenoperators of $\mathcal{L}_P$.}

The accelerated relaxation effect is at the heart of the Mpemba effect, which has been introduced and discussed in several previous works (see e.g. \cite{QM3,QM8}).  {We remark that an exponentially faster relaxation does not automatically imply the occurrence
of the Mpemba effect. It is also required that the state that decays faster is
initially farther from equilibrium than other state that decays slower. Since accelerated relaxation occurs for initial population distributions $P_n(0)$ that match the first $r$ moments of the stationary distribution $P_n^{(S)}$, such rapidly-relaxing initial states get closer and closer to $P_n^{(S)}$ as $r$ is increased, so that the emergence of the Mpemba effect should be carefully examined.} To quantify the relaxation dynamics, let $P_n^{(I)}(t)$ and $P_n^{(II)}(t)$ be the solutions to Eq.(4), corresponding to two different initial distributions of populations $P_n^{(I)}(0)$ and $P_n^{(II)}(0)$, and let $D_{P} (t)$ be a measure of the 'distance' between the distribution $P_n(t)$ and the equilibrium (thermal) distribution $P_n^{(S)}$. The definition of the distance $D_{P}(t)$
 is not  unique, and different measures have been considered in previous works, including the Hilbert-Schmidt distance, the trace distance, and the Kullback-Leibler divergence (see e.g. \cite{QM2,QM3,QM11,QM16}). 
 Here we  use the Kullback-Leibler divergence (also called relative entropy), given by \cite{QM2,QM11}
  \begin{eqnarray}
  D_{P}(t)=\sum_{n=0}^{\infty} P_n(t) \ln  \left( \frac{P_n(t)}{P_n^{(S)}} \right)
  \end{eqnarray}
 as a measure of the distance of the population distribution $P_n(t)$  from the equilibrium distribution $P_n^{(S)}$ \footnote{ We found that our results hold when considering the other two measures as well, i.e. the Hilbert-Schmidt and trace distances.}. 
   Note that $D_{P}(t)$ is non-negative and  $D_{P}(t)=0$ if and only if $P_n(t)$ is the thermal (equilibrium) distribution $P_n^{(S)}$ at temperature $T$. Note also that, if the relaxation rate of $P_n(t)$ toward the equilibrium distribution is $ |\lambda_{\alpha}|$ [the leading non-vanishing term in the series on the right hand side of  Eq.(24)], then $D_P(t) \rightarrow 0$ exponentially at a rate $2 |\lambda_{\alpha}|$. 
    The Mpemba effect arises whenever, for two assigned initial population distributions $P_n^{(I)}(0)$ and  $P_n^{(II)}(0)$ such that  $D_{P^{(II)}} (0)>D_{P^{(I)}} (0)$, asymptotically for $t \rightarrow \infty$ one has $D_{P^{(II)}} (t)<D_{P^{(I)}} (t)$. In other words, the initially state further from the equilibrium thermalizes faster, which is the manifestation of the Mpemba effect; for a sketch see Figs.1(c) and (d).\\
    To unveil the emergence of the Mpemba effect, let us consider as an example the following two initial non-equilibrium population distributions
    \begin{equation}
    P_n^{(I)}(0)=\frac{1}{1+n_{th}^{'}} \left( \frac{n_{th}^{'}}{1+n_{th}^{'}} \right)^n,
    \end{equation}
    corresponding to a thermal state but with a different temperature $T'$ (here $n_{th}^{\prime}=(\exp( \hbar \omega_0/k_BT^{\prime})-1)^{-1}$ is the mean number of quanta at temperature $T^{\prime}$), and
    \begin{equation}
    P_n^{(II)}(0)= \left\{
    \begin{array} {cc}
    p & n=n_1 \\
    1-p & n=0 \\
    0 & n \neq 0, n_1
    \end{array}
    \right.
    \end{equation}
where the integer $n_1$ is chosen such that $n_1 \geq n_{th}$ and $p=n_{th}/n_1$. Note that, when $n_{th}$ is an integer, after letting $n_1=n_{th}$ the state II corresponds to a single Fock state with excitation number $n_1=n_{th}$. The  Kullback-Leibler distances from the equilibrium distribution $P_n^{(S)}$ of such two states can be readily calculated using Eq.(29) and read
\begin{eqnarray}
D_{P^{(I)}}(0) & = & \ln \left(  \frac{1+n_{th}}{1+n_{th}^{'}} \right)+n_{th}^{'} \ln \left(  \frac{n_{th}^{'}(1+n_{th})}{n_{th}(1+n_{th}^{'})} \right)\\
D_{P^{(II)}}(0) & = & p \ln \left(  \frac{p(1+n_{th})^{n_1+1}}{n_{th}^{n_1}}  \right)+(1-p) \ln \left[(1-p)(1+n_{th})  \right].
\end{eqnarray}
Clearly, since $D_{P^{(I)}}(0)  \rightarrow 0$ as $T \rightarrow T^{\prime}$, one can choose the temperature $T^{\prime}$ close enough to $T$ such that $D_{P^{(I)}}(0)<D_{P^{(II)}}(0)$, {i.e. state II is initially farther from equilibrium than state I}. According to Eq.(26), the population distribution I has a non-vanishing spectral amplitude $C_1$ since $n^{\prime}_{th} \neq n_{th}$, so that the thermalization occurs at the rate $|\lambda_1|= 2 \gamma$, and thus $D_{P^{(I)}}(t)$ decays to zero at long times as $\sim \exp(-4 \gamma t)$. On the other hand, the initial mean number of excitations of the population distribution II is $n_{th}$, and thus according to Eq.(26) the spectral amplitude $C_1$ vanishes and the thermalization is faster and occurs with the rate $|\lambda_2|=4 \gamma$, corresponding to $D_{P^{(II)}}(t) \sim \exp(-8 \gamma t)$ at long times. In the special case where the additional condition $n_{th}+2n_{nth}^2=n_1^2p$ holds, i.e. when the first two moments of the distribution (31) match those of the equilibrium distribution, one has $C_1=C_2=0$, and thus thermalization is even faster and occurs at the rate $|\lambda_3|=6 \gamma$, corresponding to $D_{P^{(II)}}(t) \sim \exp(-12 \gamma t)$ at long times. In either cases, asymptotically for $t \rightarrow \infty$ one should have $D_{P^{(II)}} (t)<D_{P^{(I)}} (t)$, indicating the appearance of the Mpemba effect. This means that, for any initial non-equilibrium thermal distribution of populations at temperature $T^{\prime}$ arbitrarily close to the reservoir temperature $T$, one can always find a far from equilibrium population distribution that thermalizes faster. Finally, it is worth considering the limiting case of a bath at zero temperature, obtained by taking the limit $T \rightarrow 0^+$ and corresponding to $n_{th} \rightarrow 0$. According to our general analysis, since all the moments of $P_{n}^{(S)}$ vanish at zero temperature, for any non-equilibrium distribution $P_n(0)$ the condition (27) is never met, i.e. accelerated thermalization is prevented at zero temperature when considering the population dynamics. This result is not at odd with the recently predicted Mpemba effect for a photon field in contact with a zero temperature bath \cite{QM22}, since in that case coherences should be considered in the relaxation dynamics. 

 \begin{figure}[h]
\includegraphics[width=12cm]{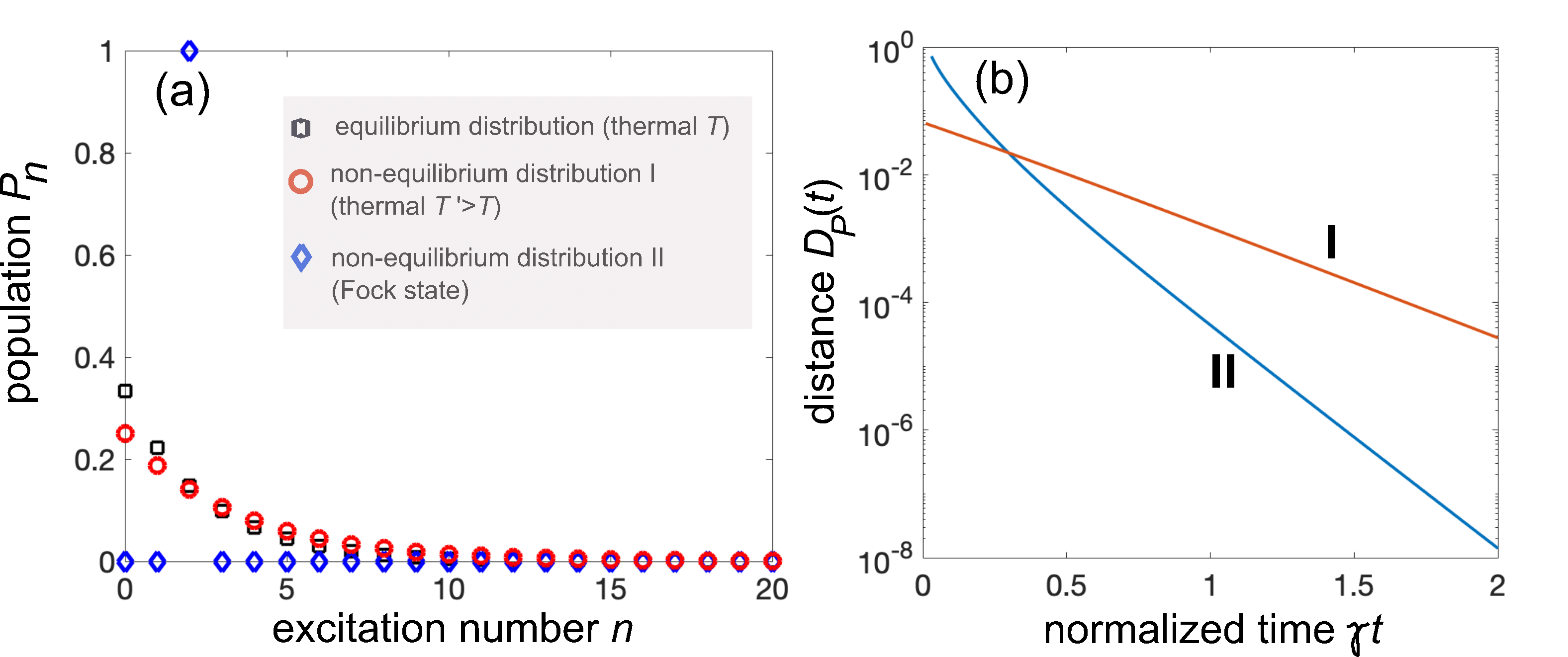}
\caption{(Color online) Mpemba effect in the relaxation dynamics of a damped quantum harmonic oscillator. (a) Behavior of the equilibrium population distribution (squares) for the quantum oscillator in contact with a reservoir at temperature $T$ such that $n_{th}=2$. The other two population distributions I and II (circles and diamonds) correspond to two initial non-equilibrium distributions, namely to a thermal distribution at temperature $T^{\prime}>T$ with $n_{th}^{\prime}=3$ (circles) and to a Fock state with excitation number $n=n_{th}=2$ (diamonds). (b) Numerically-computed behavior of the distance $D_{P}(t)$ versus normalized time $\gamma t$ for the two initial non-equilibrium distributions I and II. The crossing of the two curves at time $\gamma t \simeq 0.28$ is the signature of the Mpemba effect.}
\end{figure}
 \begin{figure}[h]
\includegraphics[width=14cm]{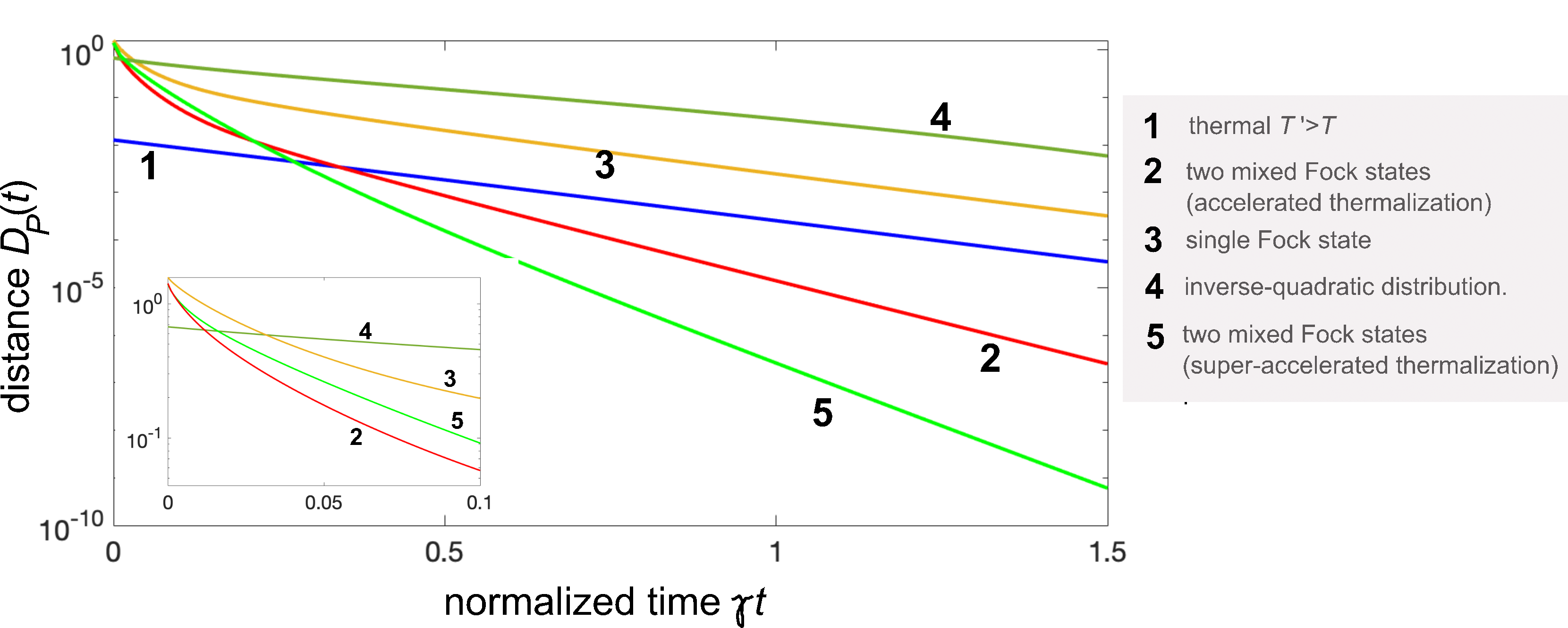}
\caption{(Color online) Thermalization dynamics (behavior of the distance $D_{P}(t)$ versus normalized time $\gamma t$ ) for a few initial out-of-equilibrium populations $P_n(0)$ in a quantum oscillator in contact with a thermal bath at temperature $T$ such that $n_{th}=2.5$. The detailed expressions of $P_n(0)$, schematically illustrated in the legend of the figure, are given in the text. {The inset at the left bottom shows an enlargement of the decay dynamics near the initial time $t=0$.}}
\end{figure}

\subsection{Numerical results}
The analytic predictions have been confirmed by numerical simulations of the Pauli master equation (8). Examples of relaxation dynamics and the occurrence of the Mpemba effect are shown in Figs.2 and 3. In the numerical simulations, the coupled equations (8) have been numerically integrated by standard matrix diagonalization methods after truncating the matrix $M$ considering  excitation numbers $n<N_{max}$, with $N_{max}=1800$. Figure 2 shows numerical results of thermalization of two initial out-of-equilibrium states I and II assuming a reservoir at temperature $T$ such that $n_{th}=2$. The population distribution I, $P_n^{(I)}(0)$, is a thermal distribution at temperature $T^{\prime}>T$ such that $n^{\prime}_{th}=3$, whereas the population distribution II, $P_n^{(I)}(0)$, is the Fock state with occupation number $n=n_{th}=2$. The behaviors of such distributions, along with the equilibrium distribution $P_{n}^{(S)}$, are shown in Fig.2(a). The relaxation dynamics toward equilibrium, shown in Fig.2(b), clearly demonstrates the  emergence of the Mpemba effect, with accelerated decay for the initial far-from-equilibrium population distribution II. Figure 3 shows the relaxation dynamics for a few initial out of equilibrium distributions 1,2,3,4, and 5 when the reservoir has a temperature $T$ such that the mean excitation number at equilibrium $n_{th}$ is not an integer ($n_{th}=2.5$). Distribution 1 is a thermal distribution at temperature $T^{\prime}>T$ such that $n_{th}^{\prime}=3$; distribution 2 is a mixture of the two Fock states $|0 \rangle$ and $|4\rangle$, as given by Eq.(31) with $n_1=4$ and $p=n_{th}/n_1=0.625$; distribution 3 is a single Fock state with excitation number $n=1$; distribution 4 is an inverse quadratic distribution that does not have finite moments, namely $P_n(0)=  (6/ \pi^2) \times 1/(1+n)^2$; finally distribution 5 is a mixture of the two Fock states $|0 \rangle$ and $|6\rangle$, as given by Eq.(31) with $n_1=6$ and $p=n_{th}/n_1=0.4167$.  Note that all distributions, except for distributions 2 and 5, do not exhibit accelerated decay as they excite the slow mode of $M$ ($C_1 \neq 0$) and the relaxation rate of $P_n(t)$ toward $P_n^{(S)}$ is $|\lambda_1|=2 \gamma$. Conversely, the initial out-of-equilibrium distribution 2 satisfies condition (27), but the second moment of $P_n(0)$ is different than the one of the equilibrium distribution $P_n^{(S)}$: in this case $C_1=0$ but $C_{2} \neq 0$, so that the thermalization process is accelerated and occurs at the rate $|\lambda_2|=4 \gamma$. Finally, for the initial out-of-equilibrium distribution 5 both first and second moments of $P_n(0)$ equal the ones of the equilibrium distribution $P_n^{(S)}$, while the third moments are different: in this case $C_1=C_2=0$ but $C_{3} \neq 0$, so that the thermalization process is super-accelerated and occurs at the rate $|\lambda_3|=6 \gamma$.

{
\section{Super-accelerated thermalization for coherences}
In the previous analysis, we demonstrated super-accelerated relaxation for certain initial states \( \hat{\rho}_i \) that are diagonal in the energy eigenbasis, where the relaxation dynamics occur entirely in the subspace \( \mathcal{L}_P \). In this case, the coherences \( \rho_{n,m} \) (\( n \neq m \)) vanish at all times. However, when the initial state contains non-vanishing coherences, the relaxation dynamics described by Eq. (7) also requires the calculation of the eigenvalues \( \lambda_k \) and the corresponding left and right eigenoperators of the Lindbladian part \( \mathcal{L}_C \).\\
Since the relaxation of coherences can be slower than that of populations, a natural question arises: Can super-accelerated relaxation also occur for initial states with non-vanishing coherences? The answer is yes. In fact, for our model, it can be shown that in the coherence subspace \( \mathcal{L}_C^{(s)} \), where \( \rho_{n,n+s} \) represents coherences with a fixed \( s \neq 0 \), the eigenvalues are given by
\begin{equation}
\lambda^{(s)}_{\alpha} = - 2\gamma \left( \alpha + \frac{|s|}{2} \right) + i s \omega_0
\end{equation}
with \( \alpha = 0, 1, 2, 3, \dots \). The technical details are provided in Appendix B.1. Equation (34) gives the complete set of eigenvalues for the coherence sub-block \( \mathcal{L}_C \) of the Lindbladian, which depend on the indices \( \alpha \) and \( s \). This set should be compared with the relaxation rates in the population sub-block, given by Eq.(20); see also Fig.1(a).
It is important to note that the two slowest decaying modes in the \( \mathcal{L}_C^{(s)} \) subspace, with a rate \( \gamma \), occur for \( s = \pm 1 \). For increasing \( |s| \), the smallest decay rate increases and is given by \( |s| \gamma \). From this result, the following property can be readily proven:\\
\\ Let us consider an initial state \( \hat{\rho}_i \), described by the density matrix \( (\hat{\rho}_i)_{n,m} = \rho_{n,m}(0) \) in the energy eigenbasis, such that the following two conditions hold:\\ 
1. The first \( r \) moments of the population distribution \( \rho_{n,n}(0) \) match those of the thermal distribution \( P_n^{(S)} \).\\
2. \( \rho_{n,m}(0) = 0 \) for \( |m - n| \leq  2r + 1 \) (\( m \neq n \)).\\
Then the state \( \hat{\rho}(t) \) exponentially relaxes toward the thermal distribution \( \hat{\rho}^{(S)} \) at a rate no slower than \( 2 \gamma (r + 1) \).\\
\\ In fact, from condition 1. and the analysis of Sec.3.3 we know that the relaxation in the population subspace   \( \mathcal{L}_P \) is accelerated and occurs at a rate no slower than \( 2 \gamma (r + 1) \).  For the coherences, \( \rho_{n,m} \) with \( |m - n|  \leq  2r + 1 \) (\( m \neq n \)) are not excited, i.e., \( \rho_{n,m}(t) = 0 \) for all subsequent times \( t \). For \( |n - m| > 2r + 1 \), the coherences decay toward zero with a rate no smaller than \( 2 \gamma (r + 1) \), according to Eq. (34). Therefore, in the series on the right-hand side of Eq. (7), the only non-vanishing terms decay to zero at a rate no slower than \( 2 \gamma (r + 1) \), ensuring that the relaxation of \( \hat{\rho}(t) \) toward the equilibrium thermal state \( \hat{\rho}^{(S)} \) is super-accelerated.\\
\\
As with the population dynamics, significant insights into the relaxation dynamics of coherences can be gained using the method of moments, which is discussed in Appendix B.2. Specifically, this method enables us to derive the following general {\em theorem}, which extends the previous result and offers a comprehensive criterion for observing super-accelerated thermalization:\\
\\
 Let us consider an initial state $\hat{\rho}_i$, described by the density matrix $(\hat{\rho}_i)_{n,m}= \rho_{n,m}(0)$ in the energy (Fock) eigenbasis, and let us assume that 
 \begin{equation}
\sum_{n=0}^{\infty}  \sqrt{\frac{(n+s)!}{n!}} \rho_{n,n+s}(0)n^l=0 
\end{equation}
for any $s=0,1,2,..$ and $l=0,1,2,...$, with $(2l+s) <h$ and $l+s>0$, where $h$ is a positive integer. Then $\hat{\rho}(t)$ exponentially relaxes toward the thermal distribution $\hat{\rho}^{(S)}$ at a rate no smaller than $ \gamma h$.\\
\\
As an illustrative example, let us  consider the relaxation of the initial state $\hat{\rho}_i= | \psi(0) \rangle \langle \psi(0)|$ corresponding to the pure state $| \psi(0) \rangle=\sqrt{1-p} |0 \rangle +\sqrt{p} |n_1 \rangle$, with $p =n_{th}/n_1 <1$, i.e.
\begin{equation}
\hat{\rho}_i= (1-p) | 0 \rangle \langle 0|+p|n_1 \rangle \langle n_1|+ \sqrt{p(1-p)} |0\rangle \langle n_1|+ \sqrt{p(1-p)} |n_1\rangle \langle 0 | .
\end{equation}
 \begin{figure}[h]
\includegraphics[width=10cm]{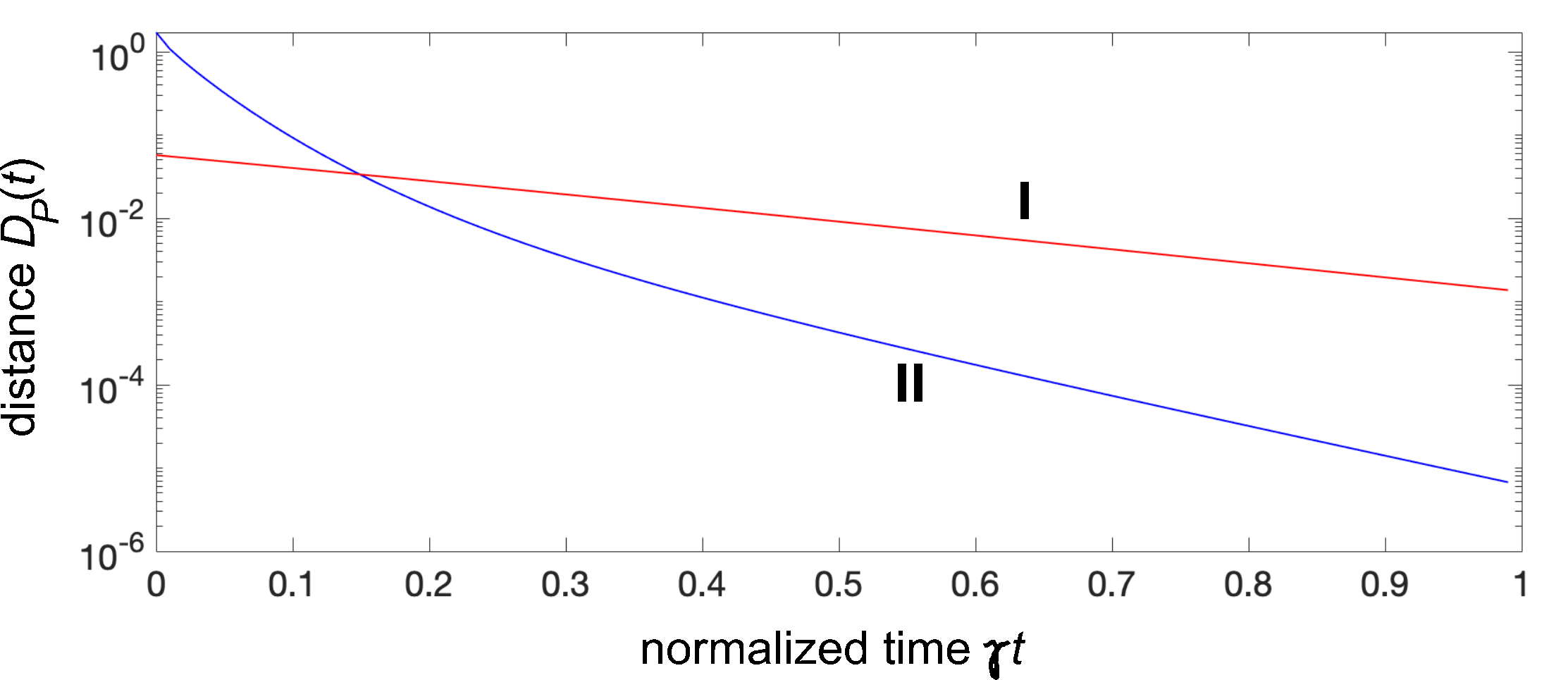}
\caption{ {(Color online) Mpemba effect in the relaxation dynamics of a damped quantum harmonic oscillator with an initial pure state with non-vanishing coherences.  The plot shows the numerically-computed behavior of the distance  $D_{\rho}(t)$ versus normalized time $\gamma t$ for two initial non-equilibrium states I and II, with a bath temperature $T$ such that $n_{th}=2$. State I is a thermal state at temperature $T^{\prime}>T$ with $n_{th}^{\prime}=3$, whereas state II is the pure state defined by Eq.(36) with $n_1=4$ and $p=n_{th}/n_1=1/2$. The crossing of the two curves  is the signature of the Mpemba effect.}}
\end{figure}
 This state differs from the mixture of the two Fock states $|0 \rangle$ and $|n_1 \rangle$, considered in Sec.3.3 [Eq.(31)], because of non-vanishing initial coherences $\rho_{0,n_1}(0)=\rho_{n_1,0}(0)= \sqrt{p(1-p)}$. Since the mean number of excitation of this state is $n_{th}$, the relaxation of the populations, $\rho_{n,n}(t)$, is accelerated and occurs at the rate $4 \gamma$.  On the other hand, the slowest rate of relaxation of coherences is given by $\gamma n_1$, according to Eq.(34) with $s=n_1$. Hence, the relaxation of the full density matrix toward the thermal equilibrium distribution occurs at a rate which is the smallest value between $4 \gamma$ and $n_1 \gamma$. For example, for $n_1 \geq 4$ the relaxation rate is $4 \gamma$. When the relaxation dynamics of such an initial pure state is compared to the one of an initial thermal state with a temperature $T'$ sufficiently close to $T$, the accelerated relaxation in the former case leads to a pronounced Mempba effect, as shown as an example in Fig.4. \footnote{{For a state $\hat{\rho}(t)$ whose density matrix $\rho_{n,m}(t)$ is not diagonal in energy eigenbasis, the Kullback-Leibler distance $D_{\rho}(t)$ between $\hat{\rho}(t)$ and $\hat{\rho}^{(S)}$ is defined as \cite{QM2} 
 $D_{\rho}(t)={\rm Tr} \left\{    \hat{\rho}(t) \left( {\rm ln} \hat{\rho}(t)- {\rm ln} \hat{\rho}^{(S)}  \right) \right\}$. Clearly, this expression reduces to Eq.(29) when $\hat{\rho}(t)$ is diagonal in the energy eigenbasis.}}  }

\section{Conclusions}
In this work, we have demonstrated {super-accelerated thermalization} and the existence of the Mpemba effect in the relaxation dynamics of a damped quantum harmonic oscillator -- a simple yet fundamental model in quantum physics. Through an analytical analysis of relaxation dynamics, we establish a clear framework for predicting and controlling thermalization of the oscillator in contact with a thermal bath. Our study reveals that the quantum harmonic oscillator can exhibit accelerated and even super-accelerated relaxation toward thermal equilibrium, an effect closely analogous to the classical Mpemba effect observed in systems such as water. Specifically, we show that any initial non-equilibrium population distribution whose first $r$ moments match those of the equilibrium distribution undergo expedited thermalization at a rate that increases linearly with $r$. Remarkably, this means that a broad class of far-from-equilibrium states can relax to equilibrium more rapidly than thermal states with initial temperatures $T^{\prime}$ arbitrarily close to the bath temperature $T$. {Super-accelerated thermalization also occurs for a broad class of initial states with non-vanishing coherences, as established by the general theorem given in Sec.4.}\\
Our findings indicate that Mpemba-like effects may be more prevalent in simple, foundational models of open quantum systems than previously recognized. This framework can be extended to more complex quantum systems and to reservoirs with customized properties, enhancing the potential applications of accelerated thermalization in quantum technologies. Rapid relaxation dynamics could enhance protocols requiring swift state resets or controlled thermalization times, making these insights potentially valuable in quantum information processing and quantum control applications.

\appendix

\section{{Relaxation dynamics in the population subspace}}
\subsection{Spectral analysis}
In this Appendix we calculate analytically the eigenvalues $\lambda$ and corresponding eigenfunctions $\psi_n$ of the following spectral problem 
\begin{equation}
E \psi_n=- \left[ n(2 n_{th}+1)+n_{th} \right] \psi_n+(n_{th}+1)(n+1)\psi_{n+1}+ n n_{th} \psi_{n-1} \equiv M \psi_n
\end{equation}
on the semi-infinite lattice $n=0,1,2,3,...$, with $\psi_n=0$ for $n<0$, {governing the relaxation dynamics in the population subspace $\mathcal{L}_P$}. In the mathematical and statistical physics literature, this spectral problem is encountered in the linear birth and death models of population dynamics and related to the properties of certain orthogonal polynomials (see e.g. \cite{birth1,birth2}).\\
Since the Jacobi matrix $M$ entering in Eq.(37) is semi-infinite, the eigenvalues and corresponding eigenfunctions do not depend solely on $M$, but also on the Hilbert space $\mathcal{H}$, i.e. the set of sequences $\{ \psi_n \}$ on which $M$ acts (see e.g. \cite{matematico}). Since we want to derive a spectral representation for the solution to the Pauli master equation (8) with arbitrary initial population distribution $P_n(0)$ with $P_n(0) \geq 0$ and $\sum_{n=0}^{\infty} P_n(0)=1< \infty$, we assume for $\mathcal{H}$ the set of sequences $\{ \psi_n \}$ such that $\sum_{n=0}^{\infty} |\psi_n| \exp( \sigma^+ n) < \infty$, where $\sigma=0^+$ is an arbitrarily small positive number.

To solve the spectral problem (37), let us introduce the generating function
\begin{equation}
G(z)=\sum_{n=0}^{\infty} \psi_n z^n,
 \end{equation}
where $z$ is a complex number with $|z|<R_0$ and $R_0$ is the radius of convergence of the series. For a sequence $\left\{ \psi_n \right\}$ in the Hilbert space $\mathcal{H}$, the series should be convergent when assuming a circle of radius $R=\exp(\sigma)=1^+$, so that $R_0>1$.
The eigenfunction $\psi_n$ is then retrieved from the inversion relation
 \begin{equation}
 \psi_n=\frac{1}{2 \pi i} \oint_{\mathcal{C}} dz G(z) z^{-n-1},
 \end{equation}
 where $\mathcal{C}$ is a closed loop in complex plane encircling counterclockwise the origin $z=0$ and internal to the convergence circle. Multiplying both sides of Eq.(37) by $z^n$ and summing up from $n=0$ to $n=\infty$, one readily obtains the following differential equation for the generating function $G(z)$
 \begin{equation}
 \frac{dG}{dz}= F(z) G(z)
 \end{equation}
 in complex $z$ plane, where we have set
 \begin{equation}
 F(z)=\frac{E+n_{th}-n_{th} z}{n_{th}z^2-(1+2n_{th})z+1+n_{th}}.
 \end{equation}
 Note that $F(z)$ has two poles at $z=z_1$ and $z=z_2$, with
 \begin{equation}
 z_1=1+1/n_{th} \; ,\;\;\; z_2=1, 
 \end{equation}
 and one can write
 \begin{equation}
 F(z)=\frac{E/n_{th}+1-z}{(z-z_1)(z-z_2)}.
 \end{equation}
Since $G(z)$ is holomorphic in the convergence domain of the series (38), for a closed loop $\mathcal{C}$ encircling $z=0$ and inside the convergence domain the integral
\begin{equation}
\oint_{C} dz \frac{1}{G} \frac{dG}{dz}= \oint_{C} d \ln G(z)
\end{equation} 
 is quantized and equal to $2 \pi i \alpha$, where $\alpha$ is an integer number. Therefore from Eqs.(40) and (43) it follows that the following quantization condition
 \begin{equation}
 \oint_{C} dz F(z)= \oint_{C} dz \frac{E/n_{th}+1-z}{(z-z_1)(z-z_2)}=2 \pi i \alpha
 \end{equation}
should be satisfied for any circle of radius $R$, with $R<R_0$. Assuming $R<1$, it can be readily shown that Eq.(45) is satisfied for any value of  $E$. Conversely, assuming $R=\exp(\sigma)=1^+$, the integral on the left hand side of Eq.(45) does depend on $E$, and thus the 'quantization condition' Eq.(45) determines the eigenvalues $E=E_{\alpha}$ of Eq.(37). Taking into account that  $z_2=1<R=1^+<z_1=R_1^2$,
 the integral on the left hand side of Eq.(45) can be  computed using the residue theorem and reads $ -2 \pi i E$, so that from the quantization condition one obtains the following set of eigenvalues 
 \begin{equation}
 E_{\alpha}=-\alpha.
 \end{equation}
 Since the eigenvalues are non-negative, $\alpha$ can takes the values $0,1,2,3,...$. For each eigenvalue $E=E_{\alpha}$, the generating function is obtained by solving Eq.(40) and reads
 \begin{equation}
 G(z)= \mathcal{N}_{\alpha} \frac{(z-z_2)^{\alpha}}{(z-z_1)^{1+\alpha}}=\mathcal{N}_{\alpha} \frac{(z-1)^{\alpha}}{(z-z_1)^{1+\alpha}}
 \end{equation}
 where $\mathcal{N}_{\alpha}$ is an arbitrary multiplication constant. The spectrum of $M$ in Hilbert space $\mathcal{H}$ is thus pure point. Note that the expression of $G(z)$ given by Eq.(47)  indicates that the radius of convergence of the series (38) is $R_0=|z_1|=1+1/{n_{th}}$, and that $\alpha$ cannot be negative since $G(z)$ should be analytic at least inside the circle of radius $R=1^+$. 
  The eigenfunction $\psi_n^{(\alpha)}$ corresponding to the eigenvalues $E_{\alpha}=-\alpha$ is finally obtained from the inversion relation (39), i.e. one has
  \begin{equation}
 \psi_n=\frac{\mathcal{N}_{\alpha}}{2 \pi i} \oint_{\mathcal{C}} dz  \frac{(z-1)^{\alpha}}{(z-z_1)^{1+\alpha}} z^{-n-1}
 \end{equation}
 \begin{figure}[h]
\includegraphics[width=7cm]{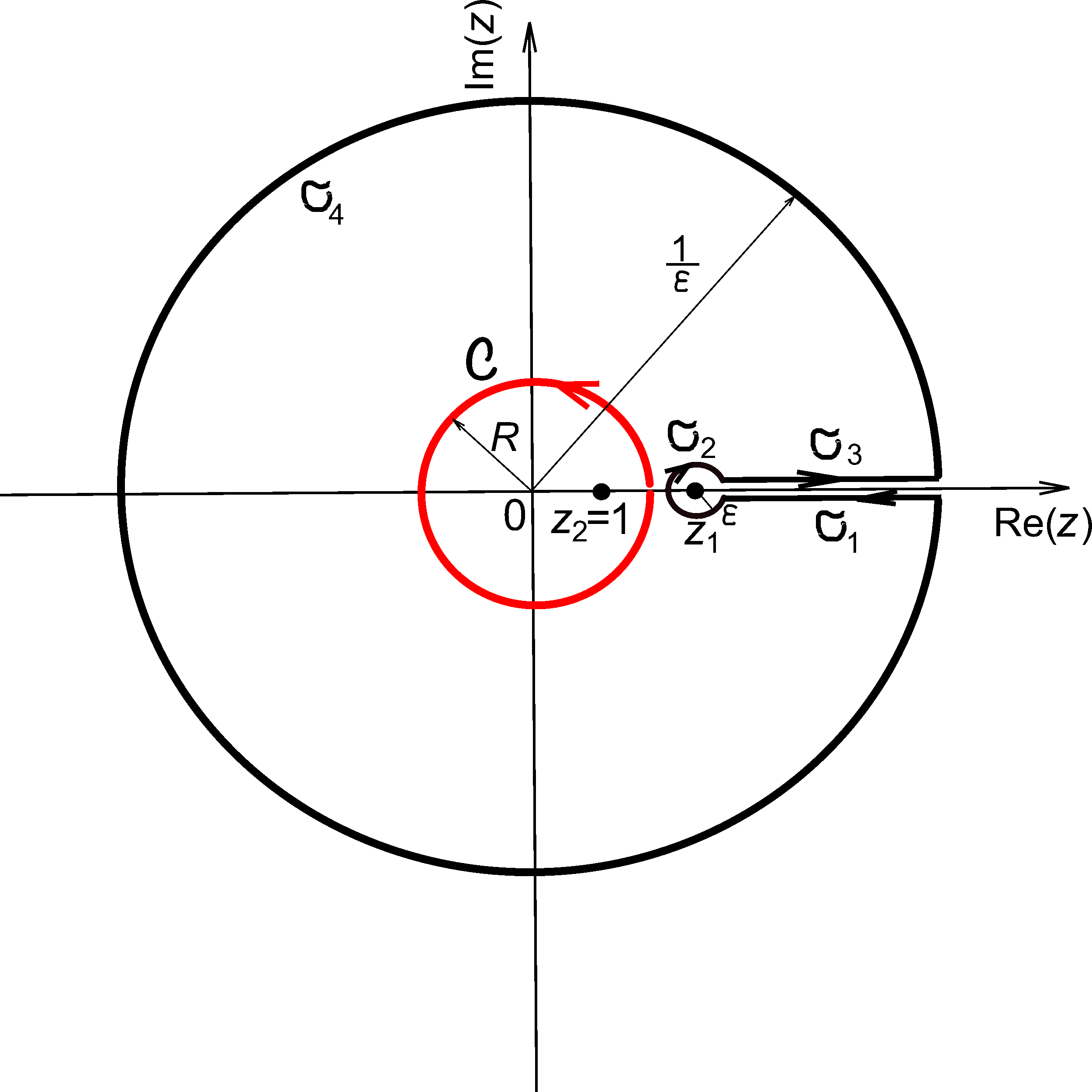}
\caption{(Color online) Contour paths in complex $z$ plane entering in the calculation of the eigenfunction amplitude $\psi_n$ via the inversion relation Eq.(48). The contour $\mathcal{C}$ is a circle of radius $R$, with $z_2=1<R<z_1=1+1/n_{th}$. The contour $\mathcal{C}$ can be deformed in the contour $\mathcal{C}^{'}=\sigma_1 \cup \sigma_2 \cup \sigma_3 \cup \sigma_4$, where $\sigma_2$ is a circle of radius $\epsilon$ and center in $z=z_1$, $\sigma_4$ is a circle of radius $1/ \epsilon$ and center in $z=0$, $\sigma_1$ and $\sigma_3$ are striaght segments close to real axis, from $z=z_1$ to $z=1/ \epsilon$. The limit $\epsilon \rightarrow 0$ should be taken.}
\end{figure}
where $\mathcal{C}$ is a circle of radius $R$, with $0<R<R_0$, traversed counterclockwise, as shown in Fig.5. Since the function under the integral on the right hand side of Eq.(48) is singular at $z=0$ and $z=z_1$, the integration contour $\mathcal{C}$ can be deformed in the closed loop $\mathcal{C}^{\prime}$ composed by the four paths $\sigma_1$, $\sigma_2$, $\sigma_3$ and $\sigma_4$, as shown in Fig.5, where $\sigma_4$ is the circle with center in $z=0$ and radius $1/ \epsilon$, $\sigma_2$ is the circle of radius $\epsilon$ centered in $z=z_1$, and $\sigma_{1,3}$ are straight lines close to the real $z$ axis and traversed in opposite directions. Hence
\begin{equation}
\oint_{\mathcal{C}}dz ...=\oint_{{\sigma_1}}dz ...+ \oint_{{\sigma_2}}dz ...+\oint_{{\sigma_3}}dz ...+\oint_{{\sigma_4}}dz ...
\end{equation}
 Clearly, since there are not branch cuts one has
 \begin{equation}
 \oint_{{\sigma_1}}dz ...+\oint_{{\sigma_3}}dz ...=0.
 \end{equation}
 Moreover, in the $\epsilon \rightarrow 0$ limit, one has $\oint_{{\sigma_4}}dz ... \rightarrow 0$, so that one obtains
 \begin{equation}
 \psi_n^{(\alpha)}= \frac{\mathcal{N}_{\alpha}}{2 \pi i} \oint_{\sigma_2} dz \frac{H(z)}{(z-z_1)^{1+ \alpha}}
 \end{equation}
 with $\epsilon \rightarrow 0$, where we have set
 \begin{equation}
 H(z)=\frac{(z-1)^{\alpha}}{z^{n+1}}.
 \end{equation}
 Since $H(z)$ is holomorphic in the neighbor of $z=z_1$, we can expand $H(z)$ is Taylor series
 \begin{equation}
 H(z)=\sum_{l=0}^{\infty} \frac{1}{l!} \left( \frac{d^l H}{dz^l} \right)_{z=z_1} (z-z_1)^l.
 \end{equation}
 After letting $z=z_1+\epsilon \exp(-i \theta)$ with $0 \leq \theta \leq 2 \pi$, from Eqs.(51) and (53) one  readily obtains
 \begin{equation}
 \psi_n^{(\alpha)}= - \frac{\mathcal{N}_{\alpha}}{2 \pi} \sum_{l=0}^{\infty}  \frac{1}{l!}\left( \frac{d^l H}{dz^l} \right)_{z=z_1}  \int_{0}^{2 \pi} d \theta \; \epsilon^{l-\alpha} \exp[-i \theta(l-\alpha)]= 
  -\mathcal{N}_{\alpha}  \frac{1}{\alpha !}\left( \frac{ d^{\alpha} H}{dz^{\alpha}} \right)_{z=z_1} ,
 \end{equation}
 i.e.
 \begin{equation}
 \psi_n^{(\alpha)}=  -\mathcal{N}_{\alpha}  \frac{1}{\alpha !}\left( \frac{ d^{\alpha}}{dz^{\alpha}} 
 \frac{(z-1)^{\alpha}}{z^{n+1}} \right)_{z=1+1/n_{th}}.
 \end{equation} 
 Without loss of generality, we may choose the constants $\mathcal{N}_{\alpha}$  independent of $\alpha$ and such that for $\alpha=0$ the eigenfunction $\psi_{n}^{(0)}$ corresponds to the thermal distribution $P_n^{(S)}$. This yields $\mathcal{N}_{\alpha}=-1/n_{th}$, so that one finally obtains
\begin{equation}
 \psi_n^{(\alpha)}= \frac{1}{ n_{th} \alpha !} \left( \frac{ d^{\alpha}}{dz^{\alpha}} 
 \frac{(z-1)^{\alpha}}{z^{n+1}} \right)_{z=1+1/n_{th}}
\end{equation} 
 which is Eq.(21) given in the main text. By calculating the derivative on the right hand side of Eq.(56) using the binomial relation, it can be readily shown that the behavior of $\psi_n^{(\alpha)}$ as $n \rightarrow \infty$ vanishes as $\sim n^{\alpha} [n_{th}/(1+n_{th})]^n$. From Eq.(12), it then follows that the left eigenfunction $\phi_n^{(\alpha)}$ is a polynomial of order $\alpha$, i.e. it is unbounded and grows algebraically as $ n \rightarrow \infty$. Finally, the following resolution of identity involving left and right eigenfunctions of $M$
 \begin{equation}
 \sum_{\alpha=0}^{\infty} \psi_n^{(\alpha)} \phi_m^{(\alpha)}= \delta_{n,m}
 \end{equation}
 holds, which justifies the spectral decomposition Eqs.(24) and (25) used in the main text. The resolution of the identity can be readily demonstrated by noting that the non-Hermitian Jacobi matrix $M$ can be transformed into a Hermitian symmetric matrix $M^{\prime}$ via a non-unitary transformation $M^{\prime}=U M U^{-1}$ with $U_{n,m}= \Delta^n \delta_{n,m}$, where $\Delta \equiv \exp(\hbar \omega_0/2 k_BT)=\sqrt{1+1/n_{th}}$. The eigenfunctions $\bar{\psi}_n^{(\alpha)}$ of the self-adjoint operator $M^{\prime}$ are given by $\bar{\psi}_n^{(\alpha)}= U \psi_n^{(\alpha)}=\psi_n^{(\alpha)} \Delta^n $, which vanish for $ n \rightarrow \infty$ as $\sim n^{\alpha} \Delta^{-n}$. Since $M^{\prime}$ is self-adjoint in $\mathcal{H}$, for the general properties of self-adjoint operators the following resolution of identity holds
 \begin{equation}
 \sum_{\alpha=0}^{\infty} \bar{\psi_n}^{(\alpha)}  \bar{\psi}_{m}^{(\alpha)}= \delta_{n,m}.
 \end{equation}
Taking into account that $\psi_n^{(\alpha)}=  \Delta^{-n} \bar{\psi}_n^{(\alpha)}$ and $\phi_n^{(\alpha)}=\Delta^{2n} \psi_n^{(\alpha)}= \Delta^n \bar{\psi}_n^{(\alpha)}$, one has
\begin{equation}
 \sum_{\alpha=0}^{\infty} \psi_n^{(\alpha)} \phi_m^{(\alpha)}= \Delta^{m-n} \sum_{\alpha=0}^{\infty} \bar{\psi}_n^{(\alpha)} \bar{\psi}_m^{(\alpha)}
\end{equation}
  and thus from Eqs.(58) and (59) it follows the resolution of identity Eq.(57).
  
  \subsection{ Method of moments}
  Main insights into the acceleration of the thermalization process can be gained by considering the temporal evolution of the moments $Q_l(t)$ of the occupation probability distribution $P_n(t)$, which are defined by
  \begin{equation}
  Q_l(t)=\sum_{n=0}^{\infty} n^l P_n(t)
  \end{equation}
with $l=0,1,2,3,...$. Clearly, $Q_0(t)=1$, $Q_1(t)$ is mean excitation number at time $t$, etc.. The values $Q_l(0)=I_l$ are determined by the initial population distribution $P_n(0)$. Multiplying both sides of Eq.(4) by $n^l$ and taking the sum over $n$ from $n=0$ to $n= \infty$, it can be readily shown that the moments $Q_l(t)$ for $l \geq 1$ satisfy the set of equations
\begin{equation}
\frac{1}{2 \gamma} \frac{dQ_l}{dt}+l Q_l= n_{th}+\sum_{k=1}^{l-1} \left\{ (-1)^{l-k-1} (1+n_{th}) \left(
\begin{array}{c}
l \\
k-1
\end{array}
\right)  +n_{th}    
\left(
\begin{array}{c}
l+1 \\
k
\end{array}
\right)
  \right\} Q_k.
\end{equation} 
Note that $Q_l(t)$ depends on the moments $Q_k$ of lower order $k<l$, so that the chain of equations (61) can be solved iteratively. In particular, for the equilibrium (thermal) distribution $P_n^{(S)}$ the moments do not vary in time and are obtained from the recursive relations 
\begin{equation}
Q_l^{(S)}=\frac{n_{th}}{l}+ \frac{1}{l} \sum_{k=1}^{l-1} \left\{ (-1)^{l-k-1} (1+n_{th}) \left(
\begin{array}{c}
l \\
k-1
\end{array}
\right)  +n_{th}    
\left(
\begin{array}{c}
l+1 \\
k
\end{array}
\right)
  \right\} Q_k^{(S)}
\end{equation}
with $Q_0^{(S)}=1$, i.e. $Q_1^{(S)}=n_{th}$, $Q_2^{(S)}=n_{th}+2n_{th}^2$ , $Q_3^{(S)}=n_{th}+67 n_{th}^2+6 n_{th}^{3}$, ...  \\
Assume now that we have a non-equilibrium initial population distribution $P_n(0)$ with 
\begin{equation}
Q_l(0)=I_l=Q_l^{(S)} 
\end{equation}
 for $l=1,2,..,r$, and 
 \begin{equation}
Q_{r+1}(0)=I_{r+1} \neq Q_{r+1}^{(S)},
\end{equation}
 i.e. for which all the first $r$ moments (with $r$ positive integer) are equal to the ones of the equilibrium thermal distribution, but not for the moment of order $(r+1)$. From Eq.(61) it readily follows that the first $r$ moments $Q_l(t)$ do not vary in time and remain locked to their equilibrium values $Q_{l}^{(S)}$, whereas the moments of order $l \geq r+1$ are non stationary and relax toward their stationary values $Q_{l}^{(S)}$ at a rate equal to $2 \gamma (r+1)$. This result, together with Eq.(60), indicates that the relaxation toward the equilibrium is accelerated whenever the first $r$ initial moments of the non-equilibrium population distribution are matched to those of the equilibrium thermal distribution, and that the relaxation occurs at the rate $2\gamma(1+r)$, which is faster as the number $r$ of equal moments increases.  In fact, let us assume that Eqs.(63) and (64) are satisfied, so that $Q_l(t)=Q_l^{(S)}(0)$ are time independent for $l=1,2,...,r$ whereas $Q_{r+1}(t)$ relaxes toward its equilibrium value at the rate $2 \gamma(1+r)$, according to Eq.(61). On the other hand, using the spectral representation (24) given in the main text one can write
 \begin{equation}
 Q_l(t)=Q_l^{(S)}+ \sum_{\alpha=1}^{\infty} C_{\alpha} \Theta_{\alpha,l} \exp(-2 \gamma \alpha t)
 \end{equation}
where we have set $\Theta_{\alpha,l} \equiv \sum_{n=0}^{\infty} n^l \psi_n^{(\alpha)}$. For consistency, from Eq.(65) it follows that the following conditions must be satisfied
\begin{eqnarray}
C_{\alpha}=0  & \alpha=1,2,..., r  \\
C_{r+1} \neq 0 \\
 \Theta_{\alpha,l}=0 & \alpha > l 
\end{eqnarray}
In particular, Eqs.(66) and (67) show that the spectral amplitudes $C_{\alpha}$ vanish for $\alpha=1,2,...,r$ while $C_{r+1} \neq 0$, so that the relaxation of $P_n(t)$ toward the equilibrium distribution is accelerated and occurs at the rate $2 \gamma(r+1)$. \footnote{Equation (68) indicates that the moments $\Theta_{\alpha,l}=\sum_{n=0}^{\infty} n^l \psi_n^{(\alpha)}$ of the $\alpha$-th eigenfunction $\psi_n^{(\alpha)}$, with order $l < \alpha$, vanish. This property can be also directly proven starting from the expression of the generating function $G(z)$ of $\psi_n^{(\alpha)}$, given by Eq.(47). In fact, after letting $z=\exp(i \theta)$, the moment $\Theta_{\alpha,l}$ is proportional to the $l$-th Taylor expansion coefficient of the function $f(\theta)=G(z=\exp(i \theta))$ near $\theta=0$, given that $f(\theta)=\sum_{n=0}^{\infty} \psi_n^{(\alpha)} \exp(i n \theta)=\sum_{n=0}^{\infty} \psi_n^{(\alpha)}[1+in \theta-(1/2) n^2 \theta^2 + ...]$ as $\theta \rightarrow 0$. Since $f(\theta)=[\exp(i \theta)-1]^{\alpha}[\exp(i \theta)-z_1]^{-(1+\alpha)} \sim (-n_{th})^{1+\alpha}  (i \theta)^{\alpha}+ o(\theta^{\alpha})$ as $ \theta \rightarrow 0$, it readily follows that the Taylor expansion coefficients  of order $l<\alpha$ vanish. }

{
\section{Relaxation dynamics in the coherence subspace}
The relaxation dynamics in the coherence subspace $\mathcal{L}_C$ is governed by the set of evolution equations for the off-diagonal elements $\rho_{n,m}(t)$ of the density matrix in the energy eigenbasis, which read [Eq.(5) in the main text]
\begin{eqnarray}
\frac{d \rho_{n,m}}{dt} & = &  2 \gamma \sqrt{(n+1)(m+1)} (1+{n}_{th}) \rho_{n+1,m+1}+2 \gamma {n}_{th} \sqrt{nm} \rho_{n-1,m-1}  \nonumber \\
 & - &  \gamma [(n+m)(1+2 {n}_{th})+2{n}_{th}] \rho_{n,m}+i \omega_0 (m-n) \rho_{n,m}.
\end{eqnarray}
Since $\rho_{n,m}(t)=\rho_{m,n}^*(t)$, we can limit our analysis considering the evolution equations for the matrix density elements with $m>n$. After letting $m=n+s$ with $s>0$ and 
\begin{equation}
\rho_{n,n+s} \equiv \sqrt{\frac{n!}{(n+s)!}} F_n^{(s)} 
\end{equation}
Eq.(69) takes the form
\begin{eqnarray}
\frac{1}{2 \gamma} \frac{dF_n^{(s)}}{dt}  & = &  (1+n_{th})(n+1)F_{n+1}^{(s)}+n_{th}(n+s) F_{n-1}^{(s)}  \nonumber \\
&  - &  \left[ n(1+2n_{th})+n_{th}+\frac{s}{2}(2n_{th}+1) \right] F_n^{(s)}+i \frac{\omega_0 s}{2 \gamma} F_n^{(s)}
\end{eqnarray}
with $n=0,1,2,3,...$. Interestingly, such equations are decoupled in $s$, indicating that the coherence subspace is block diagonal, $\mathcal{L}_C=\mathcal{L}_C^{(1)}\bigoplus \mathcal{L}_C^{(2)}\bigoplus \mathcal{L}_C^{(3)} \bigoplus ... $, where $\mathcal{L}_C^{(s)}$ is the coherence subspace defined by the off-diagonal elements $\rho_{n,n+s}$ with fixed $s$. Additionally, the form of Eq.(71) is similar to the rate equation (4) for populations, and reduces to it by letting $s=0$. Because of such a  similarity, the spectral and moment analysis methods developed in Appendix A for populations can be extended, {\it mutatis mutandis}, to the coherences in each subspace $\mathcal{L}_C^{(s)}$ with fixed index $s$. 
\subsection{Spectral analysis}
The eigenvalues $\lambda_{\alpha}^{(s)}$ in subspace $\mathcal{L}_C^{(s)}$  are obtained from the following spectral problem
\begin{equation}
E \psi_n=(1+n_{th})(n+1) \psi_{n+1}+n_{th}(n+s) \psi_{n-1}-\left[ n(2n_{th}+1)+n_{th}+\frac{s}{2}(2n_{th}+1) \right] \psi_n
\end{equation}
($n=0,1,2,3,...$), which is obtained from Eq.(71) by letting $F_n^{(s)}= \psi_n \exp(\lambda_{\alpha}^{(s) } t)$, where we have set 
\begin{equation}
E \equiv \frac{\lambda_{\alpha}^{(s)}-i \omega_0 s}{2 \gamma}.
\end{equation} 
The spectral problem defined by Eq.(72) can be analytically solved using the method of the generating function, following the same procedure as detailed in Appendix A.1. After letting $G(z)=\sum_{n=0}^{\infty} \psi_n z^n$, the following differential equation is obtained for the generating function $G(z)$
\begin{equation}
\frac{1}{G} \frac{dG}{dz}=\frac{E-n_{th}(1+s)z+n_{th}+s(n_{th}+1/2)}{n_{th}z^2-(2n_{th}+1)z+1+n_{th}} \equiv F(z).
\end{equation}
The eigenvalues $E$ are then obtained from the quantization condition  $\oint_{C} dz F(z)=2 \pi i \alpha$,
which yields
\begin{equation}
E=- \alpha-s/2.
\end{equation}
Correspondingly, one has
\begin{equation}
G(z)=\frac{(z-1)^{\alpha}}{(z-1-1/n_{th})^{1+\alpha+s}}
\end{equation}
with $ \alpha=0,1,2,3,...$. From Eqs.(73) and (75) one then obtains
\begin{equation}
\lambda_{\alpha}^{(s)}=-2 \gamma( \alpha+s/2)+i \omega_0 s.
\end{equation}
In the above analysis, we considered the relaxation dynamics of the coherences $\rho_{n,n+s}(t)$ with $s=m-n>0$. Since $\rho_{n,m}(t)=\rho_{m,n}^*(t)$, it is clear that the eigenvalues of the subset $\rho_{n,n-s}(t)$ of coherences, with $s>0$, are obtained from Eq.(77) by taking the complex conjugation. In conclusion, the whole set of eigenvalues of the coherence sub-block $\mathcal{L}_C$ of the Lindbladian is given by Eq.(34) of the main text.   
\subsection{Method of moments}
Like for the population dynamics, also for coherences the relaxation dynamics is fully embodied in the decay behavior of the moments of a certain distribution constructed from the coherences $\rho_{n,n+s}(t)$ at fixed value of $s$. To this aim, let us consider the transformation
\begin{equation}
\rho_{n,n+s}(t)=\sqrt{\frac{n!}{(n+s)!}} b_n^{(s)} \exp(i \omega_0 s t)
\end{equation}
which is basically the same as Eq.(70), but with an additional phase term that removes from the dynamics the frequency $\omega_0$. The evolution equations of the amplitudes $b_n^{(s)}$ then read
\begin{equation}
\frac{1}{2 \gamma} \frac{db_n^{(s)}}{dt}= (1+n_{th})(n+1)b_{n+1}^{(s)}+n_{th}(n+s) b_{n-1}^{(s)} 
 -   \left[ n(1+2n_{th})+n_{th}+\frac{s}{2}(2n_{th}+1) \right] b_n^{(s)}.
 \end{equation}
 Although for $s \neq 0$  $b_n^{(s)}$ does not describe any probability distribution, i.e. Eq.(79) cannot be regarded as a classical master equation like the Pauli master equation for populations, we can formally introduce the moments $Q_l^{(s)}$ via the relations
 \begin{equation}
 Q_l^{(s)} (t) \equiv \sum_{n=0}^{\infty} b_n^{(s)} (t) n^l
 \end{equation}
 where $l=0,1,2,3...$ is the moment order. Multiplying both sides of Eq.(79) by $n^l$ and taking the sum over $n$, from $n=0$ to $n=\infty$, yields the following set of dynamical equations for the moments
 \begin{eqnarray}
 \frac{1}{2 \gamma} \frac{dQ_l^{(s)}}{dt}+ \left( l+ \frac{s}{2} \right) Q_l^{(s)} & = & (1+n_{th}) \sum_{\sigma=1}^{l-1} \left(
 \begin{array}{c}
 l \\
 \sigma-1
 \end{array}
 \right) (-1)^{l-\sigma-1} Q_{\sigma}^{(s)} + n_{th} \sum_{\sigma=1}^{l-1} \left(
 \begin{array}{c}
 l \\
 \sigma-1
 \end{array}
 \right) Q_{\sigma}^{(s)}  \nonumber \\
 & + & (1+s) n_{th} \sum_{\sigma=0}^{l-1} \left(
 \begin{array}{c}
 l \\
 \sigma
 \end{array}
 \right) Q_{\sigma}^{(s)}.
 \end{eqnarray}
 Interestingly, the moment $Q_l^{(s)}(t)$ depends only on the moments $Q_k^{(s)}$ of lower order $k<l$, so that the chain of equations (81) can be solved iteratively, providing general information on  the relaxation rates of  $Q_l^{(s)}(t)$ toward the stationary state $Q_l^{(s)}(t \rightarrow \infty)=0$. In particular, if $Q_0^{(s)}(0) \neq 0$, the smallest decay rate is $\gamma s$. However, if $Q_l^{(s)}(0)= 0$ for $l=0,1,2,..,h-1$ and $Q_{h}^{(s)}(0) \neq 0$, an accelerated relaxation is obtained at the rate $2 \gamma(h+s/2)$. 
 From Eq.(80) it is clear that the slowest relaxation rate of the moments $Q_l^{(s)}(t)$, as $l$ is varied, is also the slowest decay rate of the coherences $\rho_{n,n+s}(t)$. In fact, as $l$ is varied the set of Eq.(80) can be viewed as a non-homogeneous linear system in the unknown variables $b_0^{(s)}(t)$, $b_1^{(s)}(t)$, $b_2^{(s)}(t)$,..., so that from Cramer's rule it follows that $b_n^{(s)}(t)$ is a linear superposition of the moments $Q_l^{(s)}(t)$ as $l$ is varied. Therefore, the following general theorem, stating the condition for super-accelerated thermalization of coherences, can be stated:\\
 Let us assume that the coherences $\rho_{n,m}(0)$ of the initial state $\hat{\rho}_i$ satisfy the following conditions
 \begin{equation}
 Q_l^{(s)}(0)= \sum_{n=0}^{\infty} n^l  \sqrt{ \frac{(n+s)!}{n!}} \rho_{n,n+s}(0)=0
 \end{equation}
 for any $s=1,2,...$ and $l=0,1,2,...$ with $(2l+s) < h$, where $h$ is a fixed positive integer. Then all the coherences $\rho_{n,m}(t)$ exponentially decay toward zero at a rate no smaller than $\gamma h$.\\  
 Finally, we note that the above theorem also holds for $s=0$ and $l=1,2,3,...$, i.e. for populations, reproducing the result given in Appendix A.2.
}

\end{document}